\documentclass[twocolumn,english,prb,showpacs,floatfix]{revtex4}
\usepackage{amsmath}
\usepackage{amssymb}
\usepackage{graphicx}
\usepackage{color}
\makeatletter
\@ifundefined{textcolor}{}
{%
 \definecolor{BLACK}{gray}{0}
 \definecolor{WHITE}{gray}{1}
 \definecolor{RED}{rgb}{1,0,0}
 \definecolor{GREEN}{rgb}{0,1,0}
 \definecolor{BLUE}{rgb}{0,0,1}
 \definecolor{CYAN}{cmyk}{1,0,0,0}
 \definecolor{MAGENTA}{cmyk}{0,1,0,0}
 \definecolor{YELLOW}{cmyk}{0,0,1,0}
 }

\@ifundefined{showcaptionsetup}{}{%
 \PassOptionsToPackage{caption=false}{subfig}}
\makeatother

\usepackage{babel}

\begin{document}

\title{Self-Quenching of Nuclear Spin Dynamics in Central Spin Problem}

\begin{abstract}
We consider, in the framework of the central spin $s=1/2$ model, driven dynamics of two electrons in a double quantum dot subject to hyperfine interaction with nuclear spins and 
spin-orbit coupling. The nuclear subsystem dynamically evolves in response to Landau-Zener singlet-triplet transitions of the electronic subsystem controlled by external gate voltages. Without noise and spin-orbit coupling, subsequent Landau-Zener transitions die out after about $10^4$ sweeps, the system self-quenches, and nuclear spins reach one of the numerous glassy dark states. We present an analytical model that captures this phenomenon. We also account for the multi-nuclear-specie content of the dots and numerically determine the evolution of around $10^7$ nuclear spins in up to $2\times10^5$ Landau-Zener transitions. Without spin-orbit coupling, self-quenching is robust and sets in for arbitrary ratios of the nuclear spin precession times and the waiting time between Landau-Zener sweeps as well as under moderate noise. In presence of spin-orbit coupling of a moderate magnitude, and when the waiting time is in resonance with the precession time of one of the nuclear species, the dynamical evolution of nuclear polarization results in stroboscopic screening of  spin-orbit coupling. However, small deviations from the resonance or strong spin-orbit coupling destroy this screening. We suggest that the success of the feedback loop technique for building nuclear gradients is based on the effect of spin-orbit coupling.

\end{abstract}

\author{Arne Brataas$^{1}$ and Emmanuel I. Rashba$^{2}$}
\affiliation{$^{1}$ Department of Physics, Norwegian University of Science and Technology, NO-7491 Trondheim, Norway\\
$^{2}$ Department of Physics, Harvard University, Cambridge, Massachusetts 02138,
USA}
\pacs{}

\maketitle

\section{Introduction}
\label{intro}

Electrical operation of electron spins in semiconductor double quantum dots (DQD) is one of the central avenues of semiconductor spintronics\cite{Zutic} and quantum computing.\cite{LDV,KouMar98,Levy:prl02,Awsch02} There are three basic types of electronic spin qubits, (i) the Loss-DiVincenzo\cite{LDV} qubits operating single-electron spins, 
(ii) singlet-triplet qubits operating a two-electron system and (iii) three electron qubits.\cite{DiVincenzo}  The second type 
is the center of our attention. Most widely explored singlet-triplet DQD qubits \cite{Hanson:rmp07} are based on GaAs\cite{Petta2005,Brunner} and InAs.\cite{Pfund2007,NadjPerge2010,Petersson} Both in GaAs and InAs, there are three species of nuclei possessing non-vanishing angular momenta, and the coupling between electron and nuclear spins (mostly through contact interaction) strongly influences electron-spin dynamics. Primarily, this coupling has a destructive effect causing electron spin relaxation, and many theoretical studies have focused on the challenging problem of determining the relaxation rate of an electron spin interacting with about $N\approx10^6$ nuclear spins.\cite{Merkulov:prb02,Khaetskii:prb03,Erlingsson:prb04,Coish2005,LuSham,Balents2007,Cywinski,Stanek} The problem of an electron spin interacting with a bath of nuclear spins is known as the central spin problem.

However, a controllable nuclear spin polarization, acting as an effective magnetic field, can also become a resource for manipulating electron spins.\cite{Klauser,Chekhovich} In particular, the difference (gradient) in the nuclear polarization of the left and right dots can be used for $\sigma_x$ rotations of a $S$-$T_0$ singlet-triplet qubit on the Bloch sphere.\cite{Petta2005} Efficient control of a vast ensemble of nuclear spins is very challenging, and many analytical and numerical works have been carried on this subject.\cite{Ramon:prb07,Stopa:prb10,Gullans:prl10,Rudner:prb10a,Rudner:prb10b,Burkard,Brataas:prb11,Brataas:prl12,Gullans:prb2013} The principal experimental tool for polarizing nuclear spins and building gradients is based on driving a two-electron DQD electrically through the avoided crossing ($S$-$T_+$ anticrossing) of its singlet level $S$ and the $T_+$ component of its electronic triplet $T=(T_+,T_0,T_-)$. $T_+$ is the lowest energy triplet component because the electron $g$-factor is negative, $g<0$, both in GaAs and InAs.  The width of the anticrossing is controlled by  hyperfine and spin-orbit\cite{Winkler} interactions. When the electron state changes from $S$ to $T_+$ or {\it vice versa} by passaging through the $S$-$T_+$ anticrossing, and there is no spin-orbit coupling, up to one quantum of the angular momentum is transferred to the nuclear subsystem, and such transfer facilitates polarizing the nuclear bath by performing multiple passages.

Unfortunately, experimental data show that the nuclear polarization saturates at a rather low level, typically of about 1\%.\cite{Reilly:2010} The origin of this low saturation level remains unclear and constitutes the critical obstacle for achieving higher levels of nuclear spin polarization. We recently uncovered a mechanism of dynamic self-quenching which, in absence of spin-orbit (SO) coupling, results in fast suppression of the transverse 
nuclear polarization under stationary pumping.\cite{Brataas:prl12} This is caused by screening the random field of the initial nuclear spin fluctuation by the nuclear polarization produced through pumping and closing the anticrossing. This conclusion is in a qualitative agreement with the data of Refs.~\onlinecite{Gullans:prl10,Gullans:prb2013} in the strong magnetic field limit, and the states with vanishing transverse magnetization are known as ``dark states". Meantime, by applying feedback loops, experimenters managed to achieve considerable and controllable nuclear spin polarizations.\cite{Bluhm:prl10} This poses a challenging question in which way closing the $S$-$T_+$ intersection due to the self-quenching mechanism could be avoided. Our data of Ref.~\onlinecite{Brataas:prl12} indicate that SO coupling changes the patterns of self-quenching dramatically, which implies that it is the spin-orbit coupling that might resolve the problem. The main goal of the current paper is to shed more light on the mechanisms controlling the transfer of angular momentum from the electron qubit to the nuclear bath. For this purpose, we solve numerically the equations describing coupled electron and nuclear spin dynamics for DQDs of a realistic size of more than  $N\sim10^6$ nuclear spins and a shape of two overlapping Gaussian distributions. These simulations follow up to $2\times10^5$ sweeps and unveil intimate patterns of transferring spin polarization from the electronic to the nuclear subsystem. 

To get quantitative insight onto the {\it long time} dynamics of spin pumping by multiple passages across the $S$-$T_+$ anticrossing, we restrict ourselves to the strong magnetic field regime when the Zeeman split-off $T_0$ and $T_-$ components of the electron spin triplet are well separated from the $S$ and $T_+$ states, hence, transitions to these states are disregarded. Therefore, with a semiclassical description of nuclear spins, electrons form a two-level system, and passages across the $S$-$T_+$ anticrossing are described by the Landau-Zener type theory.\cite{LL77,brevity} The detailed patterns depend on the shape on the pulses on the gates and the instantaneous nuclear configuration. In turn, during each LZ sweep the nuclear configuration changes due to the direct transfer of the angular momentum and shake-up processes.\cite{Brataas:prb11,Brataas:prl12} Between the LZ sweeps, this configuration changes because of the difference in the Larmor precession rates of different nuclear species. To follow the long term evolution, we solve the problem of the coupled dynamics self-consistently. From the mathematical point of view, we arrive to a central spin $s=1/2$ problem with a driven dynamics of the electron spin. Hence, beyond the application to the spin pumping problem, our results are of general interest for coupled dynamics of many body systems. 

In this paper we prove, both analytically and numerically, that self-quenching into dark states is a generic property of the pseudospin $s=1/2$ model in absence of SO coupling, and that self-quenching sets in after only about $10^4$ sweeps. We also demonstrate that this result stands under a moderate noise. However, the main focus of the paper is on the effect of SO coupling. Because the SO field is static while the hyperfine Overhauser field oscillates in time with the Larmor frequencies of nuclei, self-quenching cannot set in. Nevertheless, if the waiting time between LZ sweeps coincides with the Larmor period of one of the species, self-quenching sets in stroboscopically (as was demonstrated in our previous paper for a single-specie model\cite{Brataas:prl12}). More specifically, the Overhauser field of the resonant specie screens the SO field during the LZ sweeps (whose duration is small compared with Larmor periods). Therefore, during the sweeps the electron and nuclear subsystems become decoupled. As distinct from self-quenching in absence of SO coupling, the stroboscopic self-quenching is fragile. Even a small deviation from the resonance, about 1\%, destroys the delicate compensation of the SO and hyperfine contributions during the LZ sweeps. Moreover, we were able to observe the stroboscopic self-quenching only for moderate values of the SO coupling that do not exceed considerably the random fluctuations of the hyperfine field. 

We conclude that it is the SO coupling that endows the nuclear subsystem with a long term dynamics under the stationary LZ pumping. Therefore, we suggest that SO coupling is critical for efficient operating the feedback loops that require accumulation of large polarization gradients at the scale of about $10^6$ sweeps.\cite{Bluhm:prl10} Effect of SO coupling at a shorter time scale has been recently unveiled by Neder et al.\cite{neder:2013} by comparing with experimental data of Ref.~\onlinecite{Foletti:2009}.

\section{Outline and Basic Results}
\label{sec:outline}

In Sec.~\ref{sec:basic}, following two introductory sections, we present the basic equations of the driven coupled electron-nuclear spin dynamics of the central spin-$1/2$ problem that is the basis for all following calculations. In Sec.~\ref{sec:exact}, we find an analytical solution for a simple model demonstrating the phenomenon of self-quenching which reveals
basic factors controlling its rate. Our numerical technique that allows following the coupled dynamics of the electron $1/2$-pseudospin and about $10^7$ nuclear spins during up to $2\times10^5$ LZ sweeps is described in Sec.~\ref{sec:numerical}. It also includes parameters of the double quantum dot and LZ pulses used in simulations.

Sec.~\ref{sec:DNP} is the central one. It opens with the nuclear parameters of InAs and GaAs used in simulations, and includes the results of simulations and their discussion. In this section, we demonstrate that self-quenching is a generic and robust property of the coupled dynamics in absence of spin-orbit coupling, and analyze its specific features in systems consisting of two and three nuclear species. Next, we introduce SO coupling and demonstrate that it eliminates self-quenching and causes the nuclear subsystem 
to exhibit a persistent, but irregular, oscillatory dynamics. We also demonstrate the phenomenon of stroboscopic self-quenching that sets in when the waiting time between LZ sweeps is in resonance with the Larmor period of one of the nuclear species and show that it is  
very sensitive to deviations from the exact resonance. Finally, we suggest that the SO induced nuclear dynamics is critical for the feedback loop technique developed by Bluhm et al.\cite{Bluhm:prl10} for building controllable nuclear polarization gradients.

We summarize our results in Sec.~\ref{sec:conclude} and estimate the strength of SO coupling in InAs and GaAs double quantum dots in Appendix A.

\section{BASIC EQUATIONS}
\label{sec:basic}

Hyperfine electron-nuclear interactions and SO coupling govern the coupling between
electron states in $A_3B_5$ quantum dots utilized for quantum computing
purposes. Nuclear spins are dynamic and can be controlled by manipulating
magnetic fields and electronic states.

We consider electrons in double quantum dots interacting with nuclear spins via the hyperfine interaction. When there are two electrons in the dot, the orthogonal basis consists of
singlet and triplet spin states. Hyperfine and SO interactions couple these states.
By changing the gate voltages that confine electrons and determine singlet and triplet energies, a transition from a singlet $S$ to a triplet electron state $T_{+}$ (or vice versa) is accompanied by a change in the nuclear spin states. Our focus is on what happens to
the nuclear spins as we repeat Landau-Zener (LZ) transitions many times, up to $2\times10^5$, and how the changes in the nuclear spin states in turn affect electrons in the quantum dot.

We define a LZ sweep in the following way. We assume that
the quantum dot is first set in the singlet state, then a change in
the gate voltages drives a (partial) transition to the triplet state,
and finally one electron is taken out of the system and re-inserted
so that the system again is in its singlet state. During the sweep, the dynamics of the electronic qubit is controlled by the electric field produced by the gates and 
the nuclear polarization as described by Eq.~(\ref{eq:HST+}) below. In turn, semiclassical dynamics of nuclear spins is driven by the Knight fields 
 $\mbox{\boldmath$\Delta$}_{j\lambda}(t)$ arising from electron dynamics
\begin{equation}
\hbar\frac{d{\bf I}_{j\lambda}}{dt}=\mbox{\boldmath$\Delta$}_{j\lambda}\times{\bf I}_{j\lambda}\,, 
\label{dynam}
\end{equation}
where the sub index $j\lambda$ denotes a nuclear specie $\lambda$ at a lattice site $j$.  Assuming the time-scale $T_\text{LZ}$ of the LZ sweeps is much shorter than the nuclear precession times $t_\lambda$ in the external magnetic field, the total effect of the time-dependent fields  $\mbox{\boldmath$\Delta$}_{j\lambda}(t)$ on each nuclear spin can be integrated over the LZ sweep. Then the change of an individual nuclear spin during a sweep is
\begin{equation}
\triangle\mathbf{I}_{j\lambda}=\boldsymbol{\Gamma}_{j\lambda}\times\mathbf{I}_{j\lambda},
\label{eq:LZtransition}
\end{equation}
 where $\boldsymbol{\Gamma}_{j\lambda}$ accounts for the effective
magnetic field induced by the hyperfine interaction during the LZ
sweep and depends on the configuration of all the nuclear spins before the sweep. 

Landau-Zener sweeps are repeated many times. Between consecutive LZ
sweeps, electrons are in the singlet state and do not interact with
nuclear spins. During this waiting time $T_w$  between consecutive sweeps, nuclear spins precess in an external magnetic field $B$ applied along the $z$-direction. The changes of the nuclear spins between 
LZ sweeps are 
\begin{subequations}
\label{precession}
\begin{align}
\triangle I_{j\lambda}^{x} & =\cos\phi_{j\lambda}I_{j\lambda}^{x}-\sin\phi_{j}I_{j\lambda}^{y} , \\
\triangle I_{j\lambda}^{y} & =\cos\phi_{j\lambda}I_{j\lambda}^{y}+\sin\phi_{j\lambda}I_{j\lambda}^{x} , \\
\triangle I_{j\lambda}^{z} &=0, 
\end{align}
\end{subequations}
where the superscripts $x$, $y$, and $z$ denote Cartesian components of the nuclear spins, the transverse phase changes are $\phi_{j\lambda}=-2\pi T_w/t_{\lambda}$ in terms of the
spin precession times $t_{\lambda}=2\pi\hbar/g_{\lambda}\mu_{I}B$,
where $g_{\lambda}=\mu_{\lambda}/I_{\lambda}$ is the $g$-factor
for a nuclear specie $\lambda$, $\mu_{\lambda}$ is its magnetic moment,
and $\mu_{I}=3.15\times10^{-8}$ eV/T is the nuclear magneton. 

We also model the influence of noise by adding phenomenologically a random
magnetic field along the $z$-direction for each nuclear spin so that
the accumulated phases in Eq.\ \ref{precession} change 
to $\phi_{j\lambda}^\text{eff} = \phi_{j\lambda} +  \phi_{j\lambda}^\text{noise} \rightarrow-2\pi T_w\left(1/t_{\lambda}+r_{j\lambda}/\tau\right)$, where $r_{j\lambda}$ are random numbers in the interval from $-1$ to $1$. This procedure simulates a randomization of the transverse components of nuclear spins after a time of the order $\tau$,  and $\tau$ is termed the noise correlation time in what follows. While simulations described below were performed by using random sets of $r_{j\lambda}$, we mention that averaging over the noise results effectively in  changing  the phase-dependent factors in Eq.\ \ref{precession} as $\langle \cos{\phi_{j \lambda}^\text{eff} }\rangle_n=\cos{\phi_{j \lambda}} \sin{(2 \pi T_w/\tau)}/(2 \pi T_w/\tau)$, and similarly for $\langle \sin{\phi_{j \lambda}} \rangle_n$.  Therefore, this model of transverse noise leads to a semiclassical dephasing of the transverse components of nuclear spins on the time scale $\tau$.

Let us next review how electronic Landau-Zener sweeps influence nuclear spins
via $\boldsymbol{\Gamma}_{j\lambda}$.\cite{Brataas:prb11} The hyperfine
electron-nuclear interaction is
\begin{equation}
H_{hf}=V_{s}\sum_{\lambda}A_{\lambda}\sum_{j\in\lambda}\sum_{m=1,2}\delta\left(\mathbf{R}_{j\lambda}-\mathbf{r}_{m}\right)\left(\mathbf{I}_{j\lambda}\cdot\mathbf{s}(m)\right)
\label{eq:Hhf} ,
\end{equation}
 where $\mathbf{s}(m)=\boldsymbol{\mathbf{\sigma}}(m)/2$ are the
electron-spin operators in terms of the vector of Pauli matrices $\boldsymbol{\sigma}(m)$
for each electron $m=(1,2)$, $\mathbf{I}_{j\lambda}$ are the nuclear
spin operators, $A_{\lambda}$ is the electron-nuclear coupling constant
for a specie $\lambda$, and $V_{s}$ is the volume per single nuclear
spin. We consider GaAs or InAs quantum dots below; hyperfine coupling parameters for them can be found in Sec.~\ref{sec:DNP}. 

Assuming that gate voltages keep the system close to the singlet $S$ - triplet $T_{+}$ transition, the effective Hamiltonian describing the electron qubit is 
\begin{equation}
H^{(ST_{+})}=\left(\begin{array}{cc}
\epsilon_{S} & v^{+}\\
v^{-} & \epsilon_{T_{+}}-\eta
\end{array}\right),
\label{eq:HST+}
\end{equation}
 where $\epsilon_{S}$ is the singlet energy and $\epsilon_{T_{+}}$
is the triplet $T_{+}$ energy in the external magnetic field ${\bf B}=B{\hat{\bf z}}$ when nuclear spins are unpolarized. By retaining only $S$ and $T_+$ states, the problem is reduced to a 1/2 pseudospin problem, and we apply the term the central spin problem in this sense. 

The energies $\epsilon_{S}$ and $\epsilon_{T_{+}}$ are controlled by the gate voltages. The off-diagonal components $v^{\pm}=v_n^{\pm}+v_\text{SO}^{\pm}$, coupling the singlet $S$ and triplet $T_{+}$ states, contain contributions from nuclear spins 
\begin{equation}
v_n^{\pm}=V_{s}\sum_{\lambda}A_{\lambda}\sum_{j\in\lambda}\rho_{j\lambda}I^\pm_{j\lambda},\,\,
I^\pm_{j\lambda}=(I_{j\lambda}^{x}\pm iI_{j\lambda}^{y})/\sqrt{2},
\label{are equal}
\end{equation}
and SO coupling $v_\text{SO}^{\pm}$. \cite{Rudner:prb10a,Stepanenko} When  nuclear spins are polarized, the energy of the triplet state is affected by the Overhauser shift 
\begin{equation}
\eta=-V_{s}\sum_{\lambda}A_{\lambda}\sum_{j\in\lambda}\zeta_{j\lambda}I_{j\lambda}^{z} .
\label{eq6}
\end{equation}
The singlet-triplet electron-nuclear couplings are 
\begin{equation}
\rho_{j\lambda}=\int d\mathbf{r}\psi_{S}^{*}(\mathbf{r},\mathbf{R}_{j\lambda})\psi_{T}(\mathbf{r},\mathbf{R}_{j\lambda}),
\label{rho}
\end{equation}
and the electron-nuclear couplings in the $T_{+}$ state are 
\begin{equation}
\zeta_{j\lambda}=\int d\mathbf{r}\left|\psi_{T}(\mathbf{r},\mathbf{R}_{j\lambda})^{2}\right|,
\label{zeta}
\end{equation}
where $\psi_{S}$ ($\psi_{T}$) is the orbital part of the singlet
(triplet) wave function. 
Beyond the $2\times2$ $S$-$T_{+}$ model, it is possible to define a hyperfine term that determines the singlet $S$ - triplet $T_0$ coupling (as in Eq.~4 in Ref. \onlinecite{Brataas:prb11}), but it is not at the center of our attention and will not be discussed here. The effect of the electron spin $T_0$ and $T_-$ components 
is critical for the development of nuclear polarization gradients and has been investigated in Refs.~\onlinecite{Gullans:prl10,Gullans:prb2013}.

In terms of these parameters, the changes of the nuclear spins $\triangle\mathbf{I}_{j\lambda}=\boldsymbol{\Gamma}_{j\lambda}\times\mathbf{I}_{j\lambda}$
during a Landau-Zener sweep are determined by coefficients 
\begin{subequations}
\begin{align}
\Gamma_{j\lambda}^{(x)} & =-V_{s}A_{\lambda}\rho_{j\lambda}\left(Pv_{y}+Qv_{x}\right)/(2v^{2}) , \\
\Gamma_{j\lambda}^{(y)} & =V_{s}A_{\lambda}\rho_{j\lambda}\left(Pv_{x}-Qv_{y}\right)/(2v^{2}) , \\
\Gamma_{j\lambda}^{z} & =V_{s}A_{\lambda}\zeta_{j\lambda}R/(2v) ,
\end{align}
\label{eq11}
\end{subequations}
with $v^2=\left|v_{+}\right|^{2}=\left(v_{x}^{2}+v_{y}^{2}\right)/2$.
In these expressions, $0\leq P\leq1$ is the $S$-$T_{+}$ transition
probability, a real number $Q$ is the shake-up parameter defined via\cite{Brataas:prb11} 
\begin{equation}
P+iQ=-i2v^{-}\int_{-T_{\rm LZ}}^{T_{\rm LZ}}\frac{dt}{\hbar}c_{S}(t)c_{T_{+}}^{*}(t)
\label{eq:PandQ}
\end{equation}
in terms of the singlet (triplet) amplitude $c_{S}(t)$ ($c_{T}(t)$), and 
\begin{equation}
R=2v\intop_{-T_{\rm LZ}}^{T_{\rm LZ}}dt\left|c_{T_{+}}(t)\right|^{2}/\hbar
\label{R}
\end{equation}
accounts for the Overhauser shift due to the triplet $T_{+}$ component of the electron state during the interval $(-T_{\rm LZ},T_{\rm LZ})$. In the absence of SO coupling, $v_\text{SO}^\pm=0$, the change $\Delta I^z$ of the total angular momentum of nuclei $I^z=\sum_jI^z_j$ during a single sweep equals $\Delta I^z=-P$, as follows from the angular momentum conservation.\cite{Brataas:prb11}

Using the amplitudes $(c_S(t),c_{T_+}(t))$ found from solving the time-dependent Schr\"{o}dinger equation with the Hamiltonian $H^{(ST_+)}$ of Eq.~(\ref{eq:HST+}) in combination with the dynamical equations for nuclear spins of Eq.~(\ref{eq:LZtransition}) makes our approach completely self-consistent.

We assume that electrons are loaded into the singlet (0,2) state with energies far 
away from the $S$-T$_+$ anticrossing. After loading electrons, gate voltages are changed to bring the system closer to the level anticrossing, and this change is performed fast enough to keep the system in the singlet state.\cite{initial} From there on, an LZ sweep brings the system through the anticrossing. After the slow 
sweep, the system is moved back to the recharging point where it exchanges electrons with the reservoir. This back motion is fast at the scale of the narrow anticrossing and therefore does not influence the nuclear spin subsystem, but slow at the scale of the electron Zeeman splitting to keep the system inside the $S-T_+$ subspace. 

\section{Simple model with analytical solution}
\label{sec:exact}

Let us first present a simplified model that can be solved analytically and that manifests 
basic features of self-quenching\cite{Brataas:prl12} in the absence of SO coupling, $v_\text{SO}^\pm=0$. Different versions of this ``box" or ``giant spin" model were applied to various problems, see Refs.~\onlinecite{SR,DBW,Gullans:prl10,Rudner:prb10a}. Subsequently, we will in Section \ref{sec:numerical} outline a more complex and extensive numerical procedure and discuss numerical results for realistic models in Section \ref{sec:DNP}. Restricting ourselves to a single nuclear specie, we simplify the system by modelling it as a box inside which the electron wave functions are independent on position, all nuclei possess the same Larmor frequency, and all hyperfine coupling constants $A_\lambda$ and electron-nuclei coupligns $\rho_{j\lambda}$ are equal, $A_\lambda={\bar A}$ and $\rho_{j\lambda}={\bar \rho}$; typically, ${\bar A}\approx10^{-4}$ eV. Then Eqs.~(\ref{are equal}) and (\ref{rho}) simplify to $v^\alpha=A_0I^\alpha, \alpha=(+,-,z)$ with $A_0=V_s{\bar A}{\bar\rho}\sim{\bar A}/N$. Here $N$ is the number of nuclei in the box, and $I^\alpha=\sum_jI_j^\alpha$ are components of the ``giant" collective angular momentum of nuclei. In the framework of this model, nuclear spin precession in the Zeeman and Overhauser fields does not influence the coupled electron-nuclear spin dynamics, $\Delta^z=0$. With these assumptions, Eq.~(\ref{dynam}) becomes
\begin{equation}
\frac{dI^+}{dt}=-\frac{i}{\hbar}\Delta^+I^z,\,\frac{dI^-}{dt}=\frac{i}{\hbar}\Delta^-I^z,\,\frac{dI^z}{dt}=\frac{i}{\hbar}(\Delta^+I^--\Delta^-I^+).
\label{dyn}
\end{equation}
For LZ pulses, the energy level difference changes linearly with $t$, $\epsilon_S(t)-\epsilon_{T_+}(t)=\beta^2t/\hbar$, for $-T_{\rm LZ}\leq t\leq T_{\rm LZ}$, and the 
dynamics of the qubit is controlled by a dimensionless parameter $\gamma=v^+v^-/\beta^2$.   
The equation of motion for $\gamma$ following from (\ref{dyn}) is
\begin{equation}
\frac{d\gamma}{dt}=-\frac{A_0^2}{2\beta^2}\frac{d}{dt}(I^z)^2.
\label{gamdyn}
\end{equation}

During each sweep, $I^z$ changes by $\Delta I^z=-P$ and  $\gamma$ changes by $\Delta\gamma=(A_0^2/\beta^2)PI^z$. Precession of the collective nuclear spin $\bf I$ in the external magnetic field during the waiting times between sweeps changes neither $I^z$ nor $\gamma$ and is disregarded. The discrete number of sweeps $n$ can be considered
as a continous variable since the changes $\Delta \gamma$ and $\Delta I^z$ during a single sweep are small as compared to $\gamma$ and $I^z$. We then arrive at
the differential equations that determine the evolution of the $\gamma(n)$ and $I^z(n)$: 
\begin{subequations}
\begin{align}
\frac{d\gamma}{dn} &=\frac{A_0^2P}{\beta^2}I^z \, ,
\label{eq:dgamma} \\
\frac{dI^z}{dn} &=-P.
\label{eq:dIz}
\end{align}
\label{eq:n}
\end{subequations}
This central result for the simple model clarifies the different modes of self-quenching. The evolution of the LZ parameter $\gamma$ that controls the LZ probability $P$ differs in two scenarios that manifest themselves for opposite signs of $I^z$. (i) When $I^z$ is initially negative, it continues to decrease (becoming more negative) and magnitudes of both $\gamma$ and the LZ probability $P$ decrease, hence, the process slows down. Finally, self-quenching sets in exponentially, see Eq.~\ref{asympt} below. (ii) When $I^z$ is initially positive, $\gamma$ first increases so that the LZ probability $P$ becomes larger. However, since $I^z$ only can be reduced, it eventually becomes negative and self-quenching of scenario (i) sets in. So, self-quenching ultimately sets in generically independent on the original sign of $I^z$.

We can get a more detailed insight into the self-quenching dynamics by using the first integral of Eq.~\ref{gamdyn} \cite{also}
\begin{equation}
\gamma=-(A_0^2/2\beta^2)(I^z)^2+\gamma_0,
\label{gamma}
\end{equation}
where $\gamma_0$ is an integration constant that depends 
on the initial values of $\gamma$ and $I^z$, $\gamma_i$ and $I^z_i$. 
Obviously, Eq.\ \ref{gamma} dictates that $\gamma\leq\gamma_0$, and $\gamma_0 \ge 0$ because $\gamma \ge 0$ by definition. Therefore, 
\begin{equation}
I^z=\pm\sqrt{2}(\beta/A_0)\sqrt{\gamma_0-\gamma}, 
\label{Iz}
\end{equation}
and
\begin{equation}
d\gamma/dn=\pm\sqrt{2}(A_0/\beta)\sqrt{\gamma_0-\gamma}P(\gamma).
\label{dgamma}
\end{equation}

In scenario (i), when $I^z_i<0$, the minus sign should be chosen in Eqs.~\ref{Iz} and \ref{dgamma}, and $\gamma(n)$ decreases monotonically. In scenario (ii),  
when $I^z_i>0$, the dynamics first follows the plus branches of Eqs.~\ref{Iz} and \ref{dgamma}, and $\gamma(n)$ increases until it reaches its  maximum value $\gamma=\gamma_0$. At this point, $I^z(n)$ vanishes, changes sign, and continues to decrease as follows from Eq.~\ref{gamma} and Eq.\ \ref{eq:dIz} [because $P(\gamma_0)>0$]. At the same point, the signs in Eqs.~\ref{Iz} and \ref{dgamma} switch from plus to minus, and afterwards $\gamma(n)$  decreases monotonically as follows from Eq.\ \ref{eq:dgamma}.
 
The detailed asymptotic behavior of $\gamma(n)$ for $n\rightarrow\infty$ depends on $P(\gamma)$. For long LZ sweeps with $2T_{\rm LZ}\gg\hbar/v$, $P_{\rm LZ}(\gamma)=1-e^{-2\pi\gamma}\approx2\pi\gamma$, and
\begin{equation}
\gamma(n)\propto\exp[-2\pi\sqrt{2\gamma_0}(A_0/\beta)n].
\label{asympt}
\end{equation}
Equation (\ref{asympt}) describes an exponential decay with a non-universal exponent. The rate of decay increases with decreasing $\beta$, when sweeps become more adiabatic. Therefore, in absence of SO coupling the large-$n$ behavior of $\gamma(n)$ is 
exponential, and self-quenching sets on for arbitrary initial conditions. 

Let us make a rough estimate  
of the number of sweeps $n_\infty$ before self-quenching sets in based on Eq.\ \ref{asympt}. A typical original fluctuation includes $N^{1/2}$ spins, hence, $v\sim A_0\sqrt{N}$. For LZ pulses with an amplitude of about $v$ and duration of about $\hbar/v$, $\beta$ is about $\beta\sim v$. Therefore, $n_\infty\sim\beta/A_0\sim\sqrt{N}$, i.e., about the number of nuclear spins in a typical fluctuation. The dependence of $n_\infty$ on $\beta$ demonstrates the effect of the sweep duration $T_{\rm LZ}$, $n_\infty$ is smaller for longer sweeps. A similar estimate for the length  $\Delta n$ of the exponential tail in Eq.~(\ref{asympt}), with $2\pi\sqrt{2\gamma_0}\approx10$, results in $\Delta n\sim n_\infty/10$, i.e., it is shorter than $n_\infty$ by a numerical factor.

More detailed estimates for both regimes require specific assumptions about the shape and duration of sweeps. For sufficiently long sweeps, $P_{\rm LZ}(\gamma)$ can be used for $P(\gamma)$ and Eq.~(\ref{dgamma}) can be integrated. The number of sweeps $n=n(\gamma_i,\gamma_f)$, in units of $\beta/(\sqrt{2}A_0)$, between the initial $\gamma_i$ and final $\gamma_f$ values of $\gamma$ is plotted in Fig.~1 for two modes; the value of $\gamma_0$ has been chosen equal to $\gamma_0=2$. Fig.~1(a) is plotted for $I^z_i<0$ and Fig.~1(b) for $I^z_i>0$. Front sections of $n(\gamma_i,\gamma_f)$ surfaces by $\gamma_f=0$ planes demonstrate $n_\infty(\gamma_i)$, the number of sweeps before self-quenching. For $I^z_i<0$, the curve increases fast with $\gamma_i$ and reaches its maximum value at $\gamma=\gamma_0$. It is achieved at a ridge at the $n(\gamma_i, \gamma_f)$ surface that originates from the square-root singularity in the $dn/d\gamma$ dependence and is well seen in Fig.~1(a). For $I^z_i>0$, the $n_\infty(\gamma_i)$ dependence is much slower and becomes fast only near  $\gamma=\gamma_0$. In both cases, $n_\infty\sim\beta/A_0$, in agreement with the previous estimate.

Therefore, the model not only provides analytical justification of the self-quenching phenomenon found numerically in Ref.~\onlinecite{Brataas:prl12} for systems without SO coupling but relates, for single-specie systems, two modes of behavior (monotonic and nonmonotonic) to the difference in initial conditions. It is the first analytical solution of the central spin problem (i) describing dynamical evolution of a pumped system into a ``dark state"\cite{TaylorIL:PRL2003,Gullans:prl10} and (ii) establishing a connection between the initial and final states of the system.

\begin{figure}[htbp]
\includegraphics[width=\columnwidth]{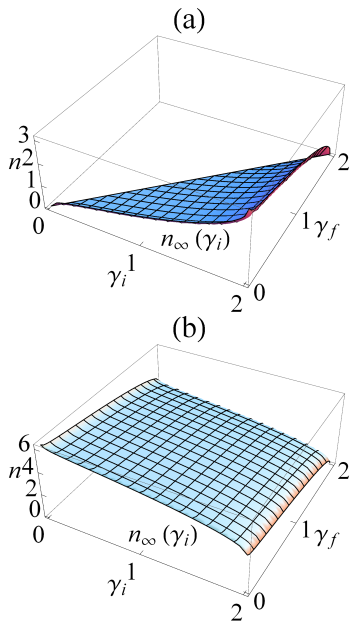}
\caption{  Number of sweeps $n$ between the initial and final values of the Landau-Zener parameter $\gamma$ for two modes; $n$ in units of $\beta/(\sqrt{2} A_0)$. (a) Initial nuclear polarization is negative, $I^z_i<0$. (b) Initial nuclear polarization is positive, $I^z_i>0$. In the plots, the lower bounds of $\gamma_i$ and $\gamma_f$ were  chosen to be  0.01 to cut  off  logarithmic singularities in $n(\gamma)$ developing because of the $P_{\rm LZ}(\gamma)$ factor in Eq.~(\ref{dgamma}). Curves $n_\infty(\gamma_i)$ in front panels show the number of sweeps before the self-quenching sets in. 
See text for details.
}
\label{Nsweepssimple}
\end{figure}

\section{Numerical Procedure}
\label{sec:numerical}

During a sweep, the difference in the singlet and triplet energies $\epsilon_{S}-\epsilon_{T_{+}}$ varies linearly in time within the sweeping interval  
$-T_{\rm LZ}\leq t\leq T_{\rm LZ}$. We impose no restrictions on the relative magnitude of the sweep duration $2T_{\rm LZ}$ and the inverse $S$-$T_+$ coupling $\hbar/v$, but, as stated above, $T_{\rm LZ}$ is long as compared to the inverse electron Zeeman energy.  Furthermore, it is assumed that the variation of the energies of both the upper and lower spectrum branches is symmetric with respect to the $S$-$T_{+}$ anti-crossing for the {\it first} transition when the initial position of the $T_{+}$ level is $\eta=\eta_{i}$. We denote the amplitude of
the change in the energy difference between the singlet and triplet
energies as $\epsilon_{\textrm{max}}$. In other words, we use 
\begin{subequations}
\begin{align}
\epsilon_{S}(t) &=\epsilon_{\textrm{max}}t/2T_{\rm LZ} \\
\epsilon_{T_{+}}(t)-\eta &=-\epsilon_{\textrm{max}}t/2T_{\rm LZ}-(\eta-\eta_{i})
\end{align}
\end{subequations}
in Eq. (\ref{eq:HST+}).\cite{numcomment} Note that as a result of the dynamical nuclear
polarization, LZ sweeps become asymmetric with respect to the anticrossing point because of the changing Overhauser shift $\eta$. There is no longer any traditional LZ passage whenever $\left|\eta-\eta_{i}\right|>\epsilon_{\textrm{max}}$, i.e. after the anticrossing point passes across one of the ends of the sweeping interval. This naturally implies a slowdown in accumulating dynamical nuclear polarization. Maintaining the LZ passages requires additional feedback mechanisms by changing the energy level difference, which we introduce below by shifting the edges of the integration interval. 

Assuming $|\eta-\eta_i|\le \epsilon_\text{max}$, the $S$ and $T_+$ states are degenerate at $t^*=-T_{\rm LZ}(\eta-\eta_i)/\epsilon_\text{max}$. To avoid trivial quenching due to the shift in $\eta$ caused by the accumulating polarization far away from the degeneracy point, the electronic energies were renormalized after every 100 sweeps keeping $\eta-\eta_i \approx 0$ and ensuring the $S$-$T_+$ anti-crossing be passed during all LZ sweeps, $-T_\text{LZ} < t^* < T_\text{LZ}$. As a result, the center of the sweep was permanently kept close to the anticrossing point. Such a regime can be achieved experimentally by applying appropriate feedback loops. 

In order to relate the properties of the sweeps to the conventional notations of the LZ transition probabilities in the limit  $T_{\rm LZ} \rightarrow \infty $, it is helpful to introduce the dimensionless initial $\tau_i=-T_{\rm LZ}[1+t^*/T_{\rm LZ}] \beta/\hbar$ 
and final $\tau_f=T_{\rm LZ}[1-t^*/T_{\rm LZ}] \beta/\hbar$ times, where $\beta=(\epsilon_\text{max} \hbar/T_{\rm LZ})^{1/2}$. The Landau-Zener parameter is $\gamma=(v/\beta)^2$. When $-\tau_i \gg \sqrt{\gamma}$ 
and $\tau_f \gg \sqrt{\gamma}$, the transition probability converges towards the Landau-Zener result $P_\text{\rm LZ} = 1 - \exp{(-2 \pi \gamma)}$. 

We consider a simple model for the electron wave functions. The orbital
part of the singlet wave function is 
\begin{align}
\psi_{S}(1,2) &=\cos\nu\psi_{R}(1)\psi_{R}(2) \nonumber \\
& +\sin\nu\left[\psi_{L}(1)\psi_{R}(2) 
+\psi_{L}(2)\psi_{R}(1)\right]/\sqrt{2}
\end{align}
and the triplet part is 
\begin{equation}
\psi_{T}(1,2)=\left[\psi_{L}(1)\psi_{R}(2)-\psi(2)\psi_{R}(1)\right]/\sqrt{2} \, ,
\end{equation}
where $\psi_{L}$ ($\psi_{R}$) denotes the wave function in the left (right) dot.  The angle $\nu$ depends on the electron Zeeman energy. We assume the electrons are in the lowest orbital harmonic oscillator state so that the wave functions are 
\begin{equation}
\psi(x,y,z)=\frac{\exp\left[-(x^{2}+y^{2})/l^{2}-z^{2}/w^{2}\right]}{\sqrt{wl^{2}(\pi/2)^{3/2}}} \, ,
\end{equation}
where $l$ is the lateral size of each dot and $w$ is its height. For two dots that are separated by a distance $d$ we form an orthonormal basis set based on the functions $\psi(x-d/2,y,z)$ and $\psi(x+d/2,y,z)$,
that defines the above $\psi_{L}$ and $\psi_{R}$. While both dots are chosen of the same size, hyperfine couplings in them differ due to the dependence of $\rho_{j\lambda}$ of Eq.~\ref{rho} on the mixing angle $\nu$.

We solve the nuclear dynamics numerically by using Mathematica 9.
First, we include all nuclear spins that are in the vicinity of the double quantum dot and satisfy the condition that the electron-nuclear coupling constants $\zeta_{j\lambda}\geq\kappa\textrm{Max}\left\{ \zeta_{j\lambda}\right\} $, where $\kappa$ is a small parameter. We checked, by changing $\kappa$, that
our results converged and have found that reducing $\kappa$ below $\kappa=0.01$ does not produce any visible changes in the plots we present. Initial configurations of the
nuclear spin directions are chosen by a pseudo-random number generator.
The initial nuclear spin configuration determines the $2\times2$ electron
$S$-$T_{+}$ Hamiltonian. We solve the time-dependent $2\times2$
differential equation numerically for linear LZ 
sweeps and compute the probability $P$, the shake-up parameter $Q$, and the
time-integrated effect of the Overhauser shift of the triplet state $T_{+}$
described by the parameter $R$. We then let the nuclear spins precess in the external magnetic field and a random noise field before the next LZ 
sweep takes place. We record all electron singlet-triplet coupling parameters
as a function of the sweep number, as well as $P$, $Q$, and the change
in the total magnetization. 

We choose realistic parameters for a double quantum dot of a height
$w=3$ \AA, size $l=50$ \AA, and distance $d=100$ \AA. We consider an external magnetic field of $B=$10 mT. Using a cut-off $\kappa=0.01$ implies that we explicitly include  in our calculations around ten millions spins. A single such calculation takes about one week on our state-of-the-art workstation. We have studied the evolution of the nuclear spin dynamics durin up to $2 \times 10^5$ LZ sweeps for $10^7$ spins and used various pseudo-random initial configurations of nuclear spins. While the detailed pattern of the dynamics depends on the initial conditions, all basic regularities were exactly the same in all simulations. Hence, our results are representative for the generic behavior of a pumped electron-nuclear system. 

\section{Dynamical nuclear polarization}
\label{sec:DNP}

We are now ready to discuss numerical results for dynamical polarization of nuclear spins. In all our simulations we consider double dots of the size $w=3$ nm, $l=50$ nm, and $d=100$ nm.

GaAs (InAs) has  8 nuclear spins per cubic unit cell so that the effective volume per site is $V_{s}=a^{3}/8$, where the lattice constant is $a=5.65$\AA\  ($a=6.06$\AA ).   When all nuclear spins are fully polarized  in GaAs (InAs), the Overhauser field seen by the electrons is $5.3$ T (0.86 T). We accept the following values of electron $g$-factors, $g_\text{GaAs}=-0.44$ ($g_\text{InAs}=-8$). The other parameters reflecting the abundance, nuclear $g$-factors, and hyperfine coupling constants are listed in Table \ref{GaAs:speciesvalues} for GaAs and Table \ref{InAs:speciesvalues} for InAs. From these values, it can be understood that in our simulations GaAs behaves as a three-specie system, whereas InAs behaves as a two-specie system. Although there are three distinct species in InAs, two of them behave in the same way with respect to the precession rate in an external magnetic field and the coupling to 
electrons so that InAs is an effective two-specie system.
\begin{table}[h]
\begin{center}
\begin{tabular}{|r|r|r|r|}
\hline
 & $^{69}$Ga & $^{71}$Ga & $^{75}$As  \\
\hline
$p$ & 30\% & 20\% & 50\% \\
\hline
$g$ & 1.3 & 1.7 & 0.96 \\
\hline
$A$ ($\mu$eV)& 77 & 99 & 94 \\
\hline
$I$ & 3/2 & 3/2 & 3/2 \\
\hline
\end{tabular}
\caption{\label{GaAs:speciesvalues} Nuclear abundances $p$, nuclear $g$-factors, hyperfine coupling constants $A$, and nuclear spin in GaAs.\cite{Schliemann:jpcm03,Taylor:prb07}}
\end{center}
\vspace{-0.6cm}
\end{table}

\begin{table}[h]
\begin{center}
\begin{tabular}{|r|r|r|r|}
\hline
 & $^{113}$In & $^{115}$In & $^{75}$As  \\
\hline
$p$ & 2\% & 48\% & 50\% \\
\hline
$g$ & 1.2 & 1.2 & 0.96 \\
\hline
$A$ ($\mu$eV)& 140 & 140 & 76 \\
\hline
$I$ & 9/2 & 9/2 & 3/2 \\
\hline
\end{tabular}
\caption{\label{InAs:speciesvalues} Nuclear abundances $p$, nuclear $g$-factors, hyperfine coupling constants $A$, and nuclear spin in InAs.\cite{Gueron:pr64,Syperek:prb11}}
\end{center}
\vspace{-0.6cm}
\end{table}

We will consider systems with different number of nuclear species to deduce coupled electron-nuclear dynamics phenomena that are robust with respect to or strongly influenced by the number of species. To this end, we choose InAs and GaAs as model systems. These systems have different magnitudes of the SO splitting;  it is modest in GaAs but strong in InAs, see Appendix \ref{sec:SO} for details. In Sec. \ref{InAs}, where calculations for nuclear parameters of InAs of Table \ref{InAs:speciesvalues} are carried out, we use modest values of SO coupling $v_\text{SO}^\pm$ to illustrate how the dynamics becomes increasingly complex and irregular with increasing strength of the spin-orbit interaction. Nevertheless, this allows making conclusions about the expected nuclear dynamics in InAs for realistic values of $v_\text{SO}^\pm$, see the end of Sec.~\ref{InAs}. The SO coupling constant $v_\text{SO}^\pm$ is a complex number. Without loss of generality, we will assume in the remainder of the paper that it is real and positive, as well as use a simplified notation, $v_\text{SO}^\pm=v_\text{SO}$.

\subsection{Two-specie systems: InAs}
\label{InAs}

Let us first consider InAs which effectively consists of two species because the parameters of $^{113}$In and $^{115}$In practically coincide. Therefore, species $^{113}$In and $^{115}$In behave as a single  specie and $^{75}$As as a second specie. We  first demonstrate that, in absence of SO coupling, the dynamical evolution of nuclear spins in InAs is similar to the dynamics in GaAs reported earlier.\cite{Brataas:prl12} In all our InAs simulations, we start in the same initial (pseudo-random) configuration of the nuclear spins. We have checked that similar results are obtained when we start in several other configurations. In all our simulations in this section, the waiting time between LZ sweeps equals the precession time of specie $^{75}$As in the external magnetic field, $T_w=t_\text{$^{75}$As}$.

\begin{figure}[htbp]
{\includegraphics[width=0.9\columnwidth]{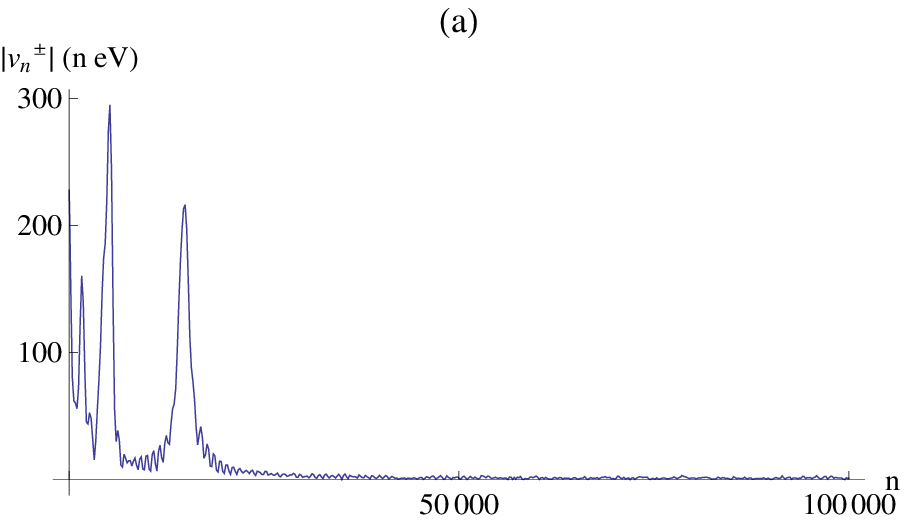}
\label{InAs1234_40n10000}} \\
{\includegraphics[width=0.9\columnwidth]{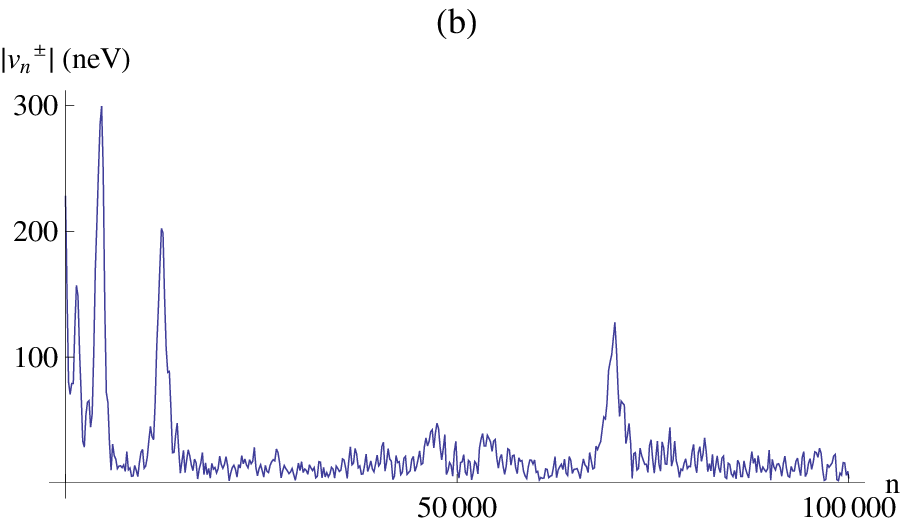} 
\label{InAs1234_50n1000}}
{\includegraphics[width=0.9\columnwidth]{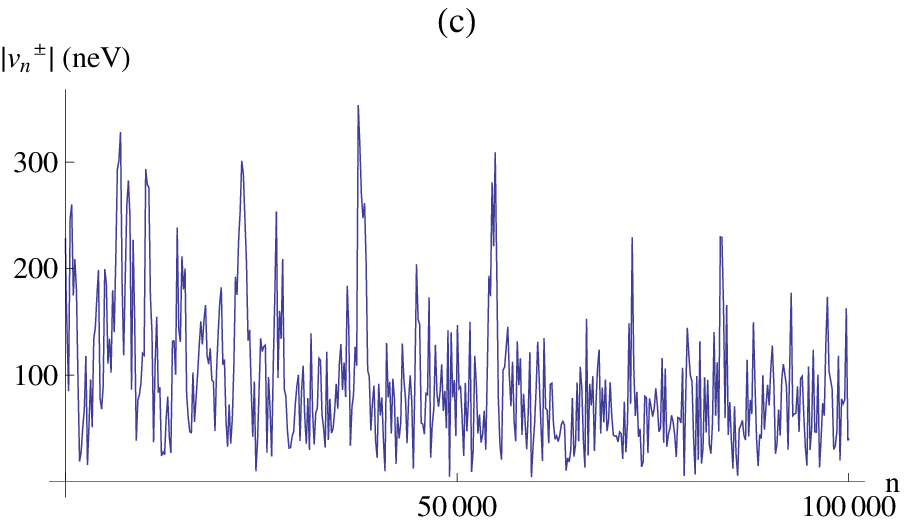} 
\label{InAs1234_50n100}}
\caption{ Hyperfine-induced singlet-triplet coupling $|v_n^\pm|$ for a double quantum dot with the nuclear parameters of InAs in absence of SO coupling with increasing level of noise as a function of sweep number $n$. (a) $\tau/t_\text{$^{75}$As}=10000$, (b) $\tau/t_\text{$^{75}$As}=1000$, and (c) $\tau/t_\text{$^{75}$As}=100$. The other parameters are $T_w=t_\text{$^{75}$As}$, $T_\text{LZ}=40$ ns.
}
\label{InAsnoise}
\end{figure}

\begin{figure}[htbp]
{\includegraphics[width=0.9\columnwidth]{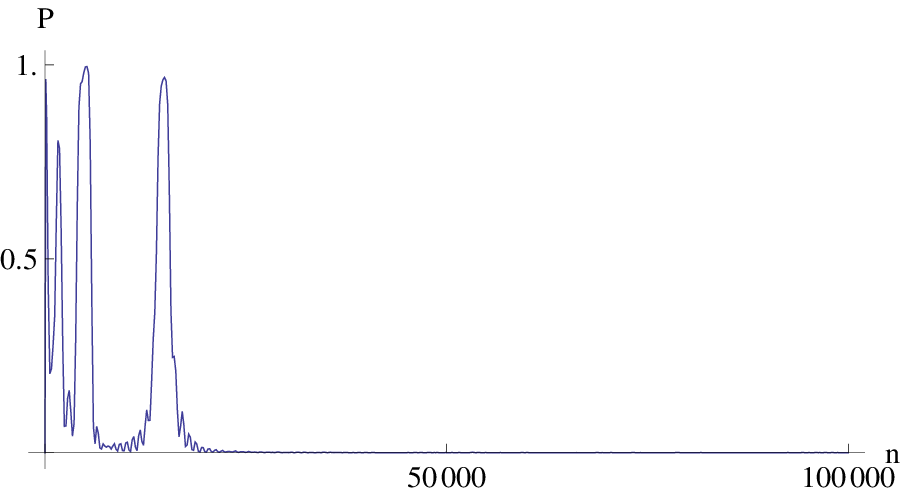}
\label{InAs1234_40n10000_P}} \\
{\includegraphics[width=0.9\columnwidth]{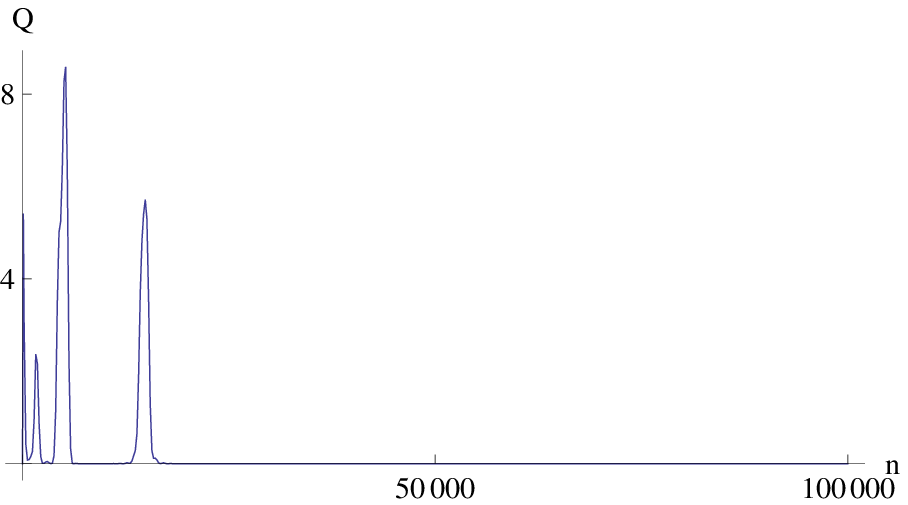} 
\label{InAs1234_50n10000_Q}}
\caption{(a)  Landau-Zener transition probability $P$ as function of the sweep number $n$ for a double quantum dot with the nuclear parameters of InAs in absence of SO coupling. (b) Shake-up parameter $Q$ as a function of sweep number $n$. The parameters are as in Fig.~\ref{InAsnoise}(a)
}
\label{InAsPandQ}
\end{figure}

We start by presenting results for a system without spin-orbit coupling, $v_\text{SO}=0$, to prove that self-quenching occurs and investigate its stability with respect to nuclear noise. Fig.\ \ref{InAsnoise}(a) shows the evolution of the magnitude of the singlet-triplet coupling $|v_n^\pm|$ with increasing number of sweeps $n$. For a sweep duration of $T_\text{LZ}=40$ ns,
the initial LZ probability for the first few sweeps is $P\sim0.5$, see Fig.~\ref{InAsPandQ}, and the singlet-triplet coupling is self-quenched already after about 20000 sweeps.  The number of sweeps $n$  required to reach self-quenching is about the same as for GaAs.\cite{Brataas:prl12} The appearance of several peaks of $v_n^\pm$ in the range of 5000-20000 sweeps before the self-quenching sets in is typical of multi-specie systems. In contrast to single-specie systems, and especially the model of Sec.~\ref{sec:exact}, in multi-specie systems final self-quenching is usually preceeded by partial self-quenchings followed by revivals. We attribute this behavior to competition between subsystems with the different Zeeman precession times. As seen in Fig.~\ref{InAsPandQ}, in each peak of $|v_n^\pm|$ the $S$-$T_+$ transition probability $P$ increases strongly, near it $I^z$ shows a step-like behavior (not shown), and accompanying peaks of $Q$ indicate massive shakeups which flop many nuclear spins per LZ sweep. The model of Sec.~\ref{sec:exact} that only deals with the total magnetization $I^z$ does not describe such events and provides a smoothened picture of the nuclear spin evolution. 

Transverse noise transforms the dynamical evolution into a dissipative one. Fig.~\ref{InAsnoise} shows the effect of the increase of the level of noise from (a) through (b) to (c). In Fig.~\ref{InAsnoise}(a), the noise correlation time is of the order of the self-quenching time $\tau/t_\text{$^{75}$As}=10000$. In this case, transverse noise only modestly perturbs the nuclear spin evolution as compared to the
non-dissipative regime (simulated and analyzed, but not shown). 
Note the presence of a long slightly visible tail with irregular oscillations along it. In contrast, Fig.~\ref{InAsnoise}(b) and (c) demonstrate that when the transverse noise correlation time $\tau$ is shorter than the typical self-quenching set-in time in un-noisy systems, self-quenching is suppressed and eventually does not happen at all; in particular, Fig.~\ref{InAsnoise}(b) demonstrates a possibility of revivals. We conclude that 
for high noise levels the chaotic evolution of the nuclear spins persists, but the magnitudes of the peaks of $|v_n^\pm|$ seem to gradually decrease in time.

Fig. \ref{InAsnoise}(a) suggests a glassy behavior of the nuclear system with an extensive manifold of dark states separated by low barriers. In the absence of noise, 
repeated LZ sweeps cause the system to end  in one of the dark states (usually after passing through several peaks of $|v_n^\pm|$). Weak noise produces slow diffusion between adjacent dark states across low saddle points. During this diffusion, the magnetization $I^z$ changes only slightly. With increasing noise, the system experiences revivals as seen in 
Fig.~\ref{InAsnoise}(b) as a sharp peak in $|v_n^\pm|$. During such events $P$ increases strongly, $I^z$ shows step-like behavior, and peaks in $Q$ (not shown) indicate massive shakeups, similarly to the patterns discussed as applied to Fig.~\ref{InAsnoise}(a) above.

\begin{figure}[htbp]
{\includegraphics[width=0.9\columnwidth]{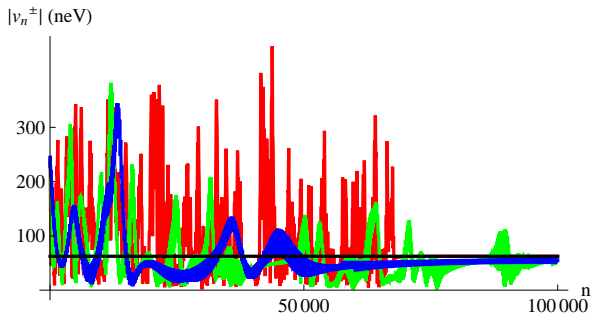}
\label{InAs1234_40n10000}} \\
\caption{ Transverse nuclear polarization for a 
double quantum dot with the nuclear parameters of InAs as a function of sweep number $n$ for the spin-orbit coupling $v_\text{SO}=62$ neV (black line) and the Landau-Zener sweep durations (red curve) $T_\text{LZ}=80$ ns, (green curve) $T_\text{LZ}=40$ ns, and (blue curve) $T_\text{LZ}=20$ ns. Resonant pumping with $T_w=t_{^{75}{\rm As}}$, the polarization is plotted at multiples of $T_w$. See text for details. 
}
\label{InAsSO}
\end{figure}

We demonstrated earlier that SO coupling is screened stroboscopically in a single-specie system.\cite{Brataas:prl12}  Next, we will demonstrate that SO coupling can be screened stroboscopically also in multi-specie systems, and investigate this phenomenon in more detail. Fig.\ \ref{InAsSO} shows simulations of the singlet-triplet coupling $v_n^\pm$ for 
$v_\text{SO}=62$ neV and three LZ sweep durations $T_\text{LZ}$. In comparison, the straight black line indicates the value of the spin-orbit coupling $v_\text{SO}=62$ neV
(which is independent of the sweep number $n$). We see that in all these simulations, the spin-orbit coupling eventually becomes screened so that all the colored lines approach the black line which implies that $|v_n^\pm|=|v_\text{SO}|$. For longer LZ sweep durations, oscillations of $v_n^\pm$ are more rapid,  but screening eventually occurs faster because nuclear spins are more strongly affected during each sweep. 

Screening of the SO coupling even in multi-specie systems sounds counter-intuitive at first glance. Indeed,  the spin-orbit coupling $v_\text{SO}$ is static while the transverse components of the nuclear spins contributing to $v_n^\pm$ precess in time. In InAs, 
two nuclear species $^{113}$In and $^{115}$In precess at the same frequency and behave effectively as a single spin specie whereas the third spin specie, $^{75}$As, precesses at a different frequency. So, while screening indicates that the magnitude of the singlet-triplet coupling $v_n^\pm$ remains finite, it must inevitably precess in time. Therefore, it cannot compensate the spin-orbit coupling $v_\text{SO}$ at all instants of time. The screening we observe is only possible because the waiting time $T_w$ is exactly equal to the precession time of the $^{75}$As specie, $T_w=t_{^{75}{\rm As}}$. The data used in our plots of $\vert v_n^\pm\vert$ were taken at exact multiples of the the waiting time, which 
was equal to the precession time of the $^{75}$As specie. Therefore, the self-quenching that manifests itself in Fig.~\ref{InAsSO} is a {\it stroboscopic self-quenching.}

Stroboscopic self-quenching can be understood in the following way. The dynamical evolution of nuclear spins causes self-quenching of the sum of the contributions from the transverse components of species $^{113}$In and $^{115}$In (that are out of resonance with the pumping period $T_w$, hence, their contribution to $v_n^\pm$ vanishes). In contrast, the contribution from the specie $^{75}$As to $v_n^\pm$ exactly compensates the spin-orbit coupling $v_\text{SO}^\pm$ at every time instant when a LZ sweep happens. In other words, the matrix elements $v_n^\pm(t)$ change in time harmonically with the amplitude $v_\text{SO}^\pm$ and a period $t_{^{75}{\rm As}}$: 
\begin{equation}
v_n^\pm(t) = v_\text{SO}^\pm \cos{(2 \pi t/t_\text{$^{75}$As})}.
\label{eq:SOscreening}
\end{equation}
This generalizes our previous findings of the screening of SO coupling in single-specie systems.\cite{Brataas:prl12} For a single-specie imitation of GaAs, we found that the SO coupling was screened in such a way that that the matrix element changed harmonically with the amplitude $v_\text{SO}$ and a period $t_\text{GaAs}$, where $t_\text{GaAs}$ is the average precession time of the three nuclear spin species in GaAs.\cite{Brataas:prl12}

\begin{figure}[htbp]
{\includegraphics[width=0.9\columnwidth]{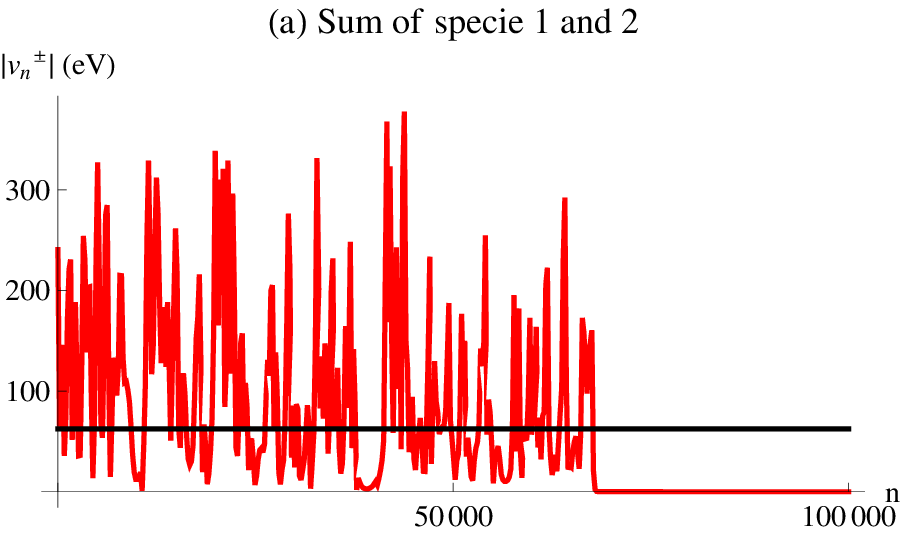}} \\
{\includegraphics[width=0.9\columnwidth]{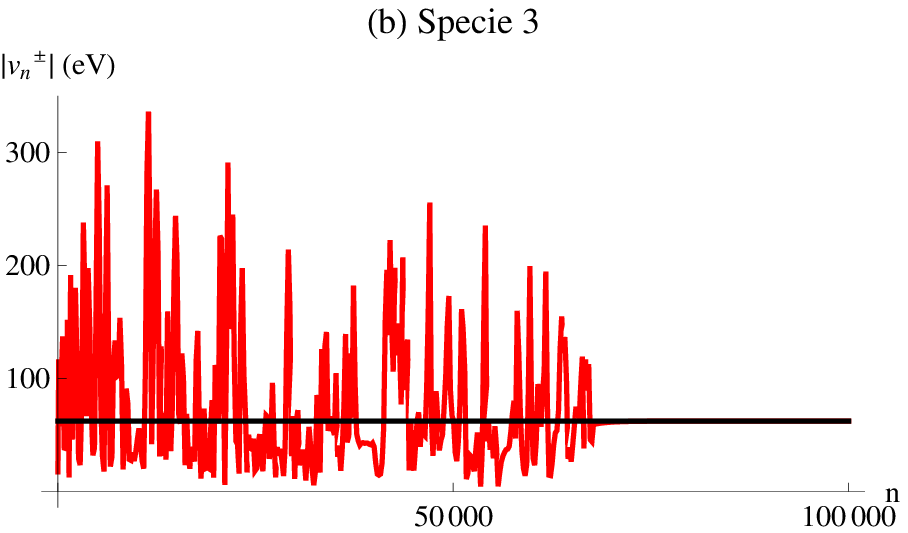}}
\caption{ (a) Sum of contributions from $^{113}$In and $^{115}$In to hyperfine-induced singlet-triplet coupling $v_n^\pm$ for spin-orbit coupling 
$v_\text{SO}=62$ neV (black line) as a function of sweep number $n$. 
 (b) Contribution from $^{75}$As to hyperfine-induced singlet-triplet coupling $v_n^\pm$ for spin-orbit coupling $v_\text{SO}=62$ neV  (black line) 
as a function of sweep number $n$. The LZ sweep duration is $T_\text{LZ}=80$ ns.
}
\label{fig:113Inand115Inand75As}
\end{figure}

Let us now demonstrate explicitly that when self-quenching sets in, the sum of the contributions from the transverse components of $^{113}$In and $^{115}$In to $|v_n^\pm|$ vanishes while the contribution from $^{75}$As equals $v_\text{SO}$. We show in Fig.~\ref{fig:113Inand115Inand75As}(a) the contribution from $^{113}$In and $^{114}$In to $|v_\text{n}^\pm|$ as a function of the number of sweeps $n$. Clearly, it vanishes for large $n$. On the other hand, $^{75}$As whose nuclear precession time equals the waiting time $T_w$, makes a contribution to $|v_n^\pm|$ that exactly compensates $|v_\text{SO}^\pm|$ 
at all integers of $T_w$, see Fig.~\ref{fig:113Inand115Inand75As}(b). Hence, Eq.~(\ref{eq:SOscreening}) is satisfied. 

We note that the contributions from $^{113}$In and $^{115}$In to $v_n^\pm$ vanish not only stroboscopically but identically, at each instant of time (not shown). We have also checked that in systems without spin-orbit coupling, self-quenching sets in for all species and for an arbitrary ratio between $T_w$ and the precession times of the species  (not shown). For three-specie systems the last statement is proven below, see Sec.~\ref{sec:GaAs}.

\begin{figure}[htbp]
{\includegraphics[width=0.9\columnwidth]{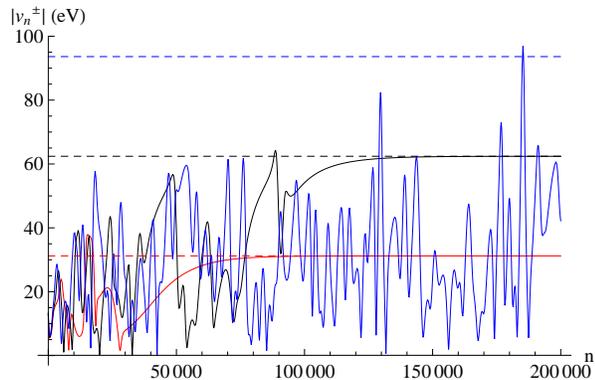}}
\caption{ Contribution of the specie $^{75}$As to the singlet-triplet coupling $v_n^\pm$ as a function of sweep number $n$ for different values of $v_\text{SO}$: 31 neV (red curve), 62 neV (black curve), and 91 neV (blue curve). Duration of LZ pulses $T_\text{LZ}=40$ ns. Waiting time between consecutive LZ pulses equals the precession time of $^{75}$As, $T_w=t_\text{$^{75}$As}$. Nuclear parameters of InAs.
}
\label{fig:InAslargeSO}
\end{figure}

Now we will illustrate that stroboscopic screening of SO coupling can be practically achieved only for small and moderate magnitudes of $v_\text{SO}$. Since stroboscopic screening implies that the contribution from $^{75}$As to $v_n^\pm$ compensates $v_{\rm SO}$ while the combined contribution from species $^{113}$In and $^{115}$In vanishes, we show in Fig.~\ref{fig:InAslargeSO} the evolution of the contribution of $^{75}$As to  $v_n^\pm$ as a function of sweep number for three values of spin-orbit coupling $v_{SO}= 31, 62$ and 91 neV.\cite{Q} While all the results in Fig.~\ref{fig:InAslargeSO} were found for the same value of $T_{\rm LZ}$ and the same initial conditions, screening sets in at $n\approx 75000$ for $v_\text{SO}=31$ neV, is delayed to $n\approx 125 000$ for $v_\text{SO}=62$neV, and is far from complete even at $n=200 000$ for $v_\text{SO}=93$ neV. These data suggest that stroboscopic self-quenching sets in when $v_{\rm SO}\alt v_n^0$, where $v_n^0\approx A/\sqrt{N}$ is a typical fluctuation of the Overhauser field, and cannot be practically achieved for $v_{SO}\agt v_n^0$; see  estimates of the magnitude of the spin-orbit coupling in 
Appendix A. This criterion resembles the criterion of the phase transition of Ref.~\onlinecite{Rudner:prb10a}.

One should keep in mind that with the interval between LZ pulses of about $1 \mu$s, a set of $n\sim10^6$ pulses takes about 1 s which is a typical scale of nuclear spin diffusion\cite{Chechnovich}, which is not taken into account in the above considerations. We expect, but have not checked, that inhomogeneity of magnetic field should have a detrimental effect on stroboscopic self-quenching. Therefore, we conclude that stroboscopic quenching of SO coupling is less generic and more fragile than self-quenching in systems without spin-orbit coupling. 

We estimate in Appendix \ref{sec:SO} that SO coupling is weaker or comparable to (stronger than) the typical nuclear polarization induced singlet-triplet coupling in GaAs (InAs). As a consequence, we expect the SO coupling might be stroboscopically screened in GaAs systems, but that stroboscopic screening is improbable in InAs systems. 

\subsection{Three-specie systems: GaAs}
\label{sec:GaAs}

In this section, we present new results for GaAs that complete the picture of the generic nature of self-quenching in multi-specie systems. Furthermore, we show that screening of the SO coupling requires that the waiting time $T_w$ is in resonance with the precession time of one of the nuclear species. When the resonance condition is not satisfied, screening of the SO coupling is partial and irregular.

In Sec.~\ref{InAs} and in Ref.~\onlinecite{Brataas:prl12}, self-quenching in multi-specie systems in absence of SO coupling was demonstarted only under the conditions when the waiting time $T_w$ was in resonance with the precession time $t_{^{75}{\rm As}}$ of the $^{75}$As specie, $T_w=t_{^{75}{\rm As}}$. We demonstrate here that while self-quenching is generic and  independent of the waiting time, the evolution towards the self-quenched states
depends on the waiting time.

\begin{figure}[htbp]
{\includegraphics[width=0.9\columnwidth]{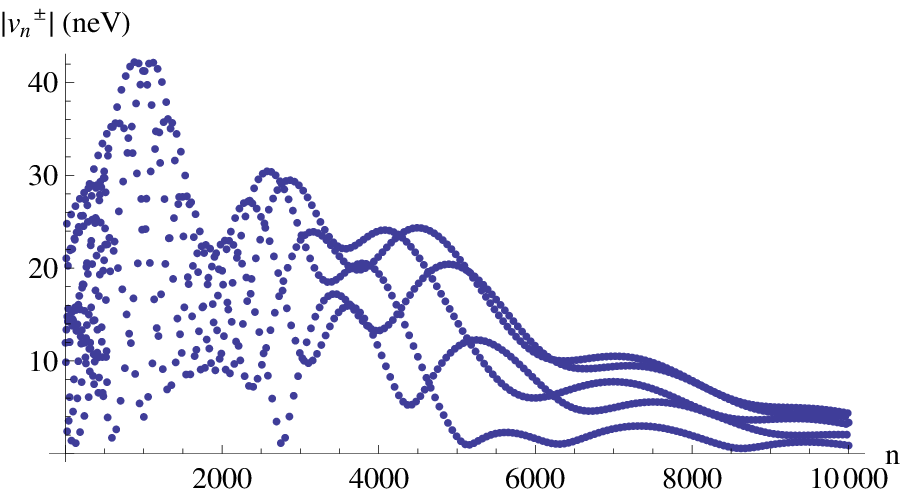}
\label{fig:GaAs_resonant}} \\
{\includegraphics[width=0.9\columnwidth]{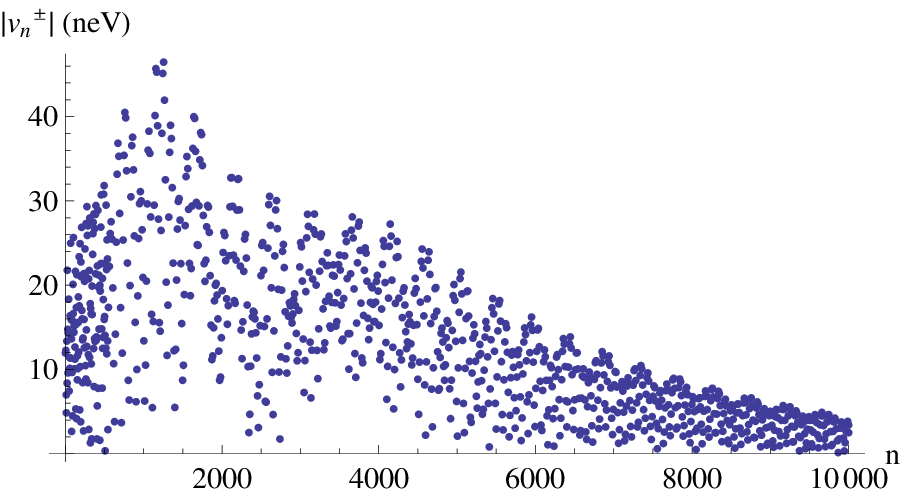}
\label{fig:GaAs_nonresonant}} \\
\caption{ Transverse nuclear polarization as a function of sweep number $n$ for a GaAs double quantum dot in absence of spin-orbit coupling and 
transverse noise. Duration of LZ pulses $T_\text{LZ}=80$ ns. (a) The waiting time is in resonance with the $^{75}$As precession time, $T_w=t_\text{$^{75}$As}=13.7 \mu s$. (b) The waiting time is incommensurate with the $^{75}$As precession time, $T_w=1.39~ t_\text{$^{75}$As}=19.1 \mu s$.
}
\label{InAstwowaittimes}
\end{figure}

To this end, we plot in Fig.~\ref{InAstwowaittimes} the evolution of the singlet-triplet coupling $v_n^\pm$ for two different values of the waiting time $T_w$. Fig.~\ref{InAstwowaittimes}(a) displays results of simulations for the resonant case when $T_w=t_{^{75}{\rm As}}$, in which pronounced oscillations are distinctly seen. For $n\agt3000$, the plot consists of five branches that reflect coupled dynamics of three species. In contrast, in the absence of the resonance, Fig.~\ref{InAstwowaittimes}(b), the evolution is chaotic. Nevertheless, self-quenching sets-in in both cases and, what is most remarkable, at the same time scale of $n\approx10^4$. Remarkably, the processes of Figs.~\ref{InAstwowaittimes}(a) and \ref{InAstwowaittimes}(b) ended in states with the same $I^z$ (not shown). While the set of dark states is vast (as follows from our discussion in Section \ref{sec:exact}), this observation indicates that the number of strong attraction centers in which self-quenching ends is more scant.

We conclude that self-quenching in systems without spin-orbit coupling is generic and robust, at least in the framework of $S$-$T_+$ scheme. 

We checked that not only does the total matrix element $v_n^\pm$ vanish, but also the matrix elements for 
all of three species contributing to it. Because between the LZ sweeps the electron subsystem is in its singlet state, the Knight shift vanishes, and according to Eq.~(\ref{precession}) all nuclei belonging to some specie precess with the same speed. Therefore, the self-organization of the nuclear subsystem that annihilates its coupling to the electron spin persists during the free precession periods.  

Finally, we demonstate that while self-quenching is a generic feature in the absence of SO coupling regardless of the ratio between the waiting time between the LZ sweeps $T_w$ and the nuclear precession times $t_\lambda$, in presence of SO coupling the stroboscopic self-quenching is not generic and highly sensitive to this ratio. Only modest deviations from the resonance destroys the screening of SO coupling. 

\begin{figure}[htbp]
{\includegraphics[width=0.9\columnwidth]{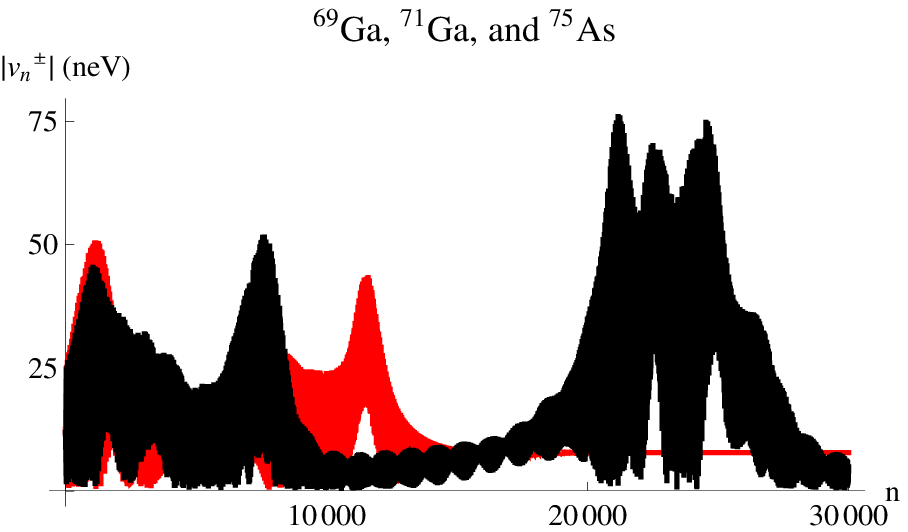}}
\caption{ Transverse nuclear polarization as a function of sweep number $n$ for a GaAs double quantum dot with $v_\text{SO}=$8 neV. The duration of LZ pulses $T_\text{LZ}=80$ ns. The red curve shows results for the waiting time in resonance with the precession time of $^{75}$As, $T_w=t_{^{75}{\rm Ar}}$, and the black curve for a 1\% deviation from the resonance. 
}
\label{fig:GaAs_SOresandnonrestot}
\end{figure}

In Fig.~\ref{fig:GaAs_SOresandnonrestot}, we plot $|v_n^\pm|$ under the conditions when the waiting time is in exact resonance with the precession time of $^{75}$As (red curve), and when there is a 1$\%$ deviation from the resonance (black curve). While the SO coupling is clearly screened in resonance, only a tiny deviation from resonance destroys screening.

\begin{figure}[htbp]
{\includegraphics[width=0.9\columnwidth]{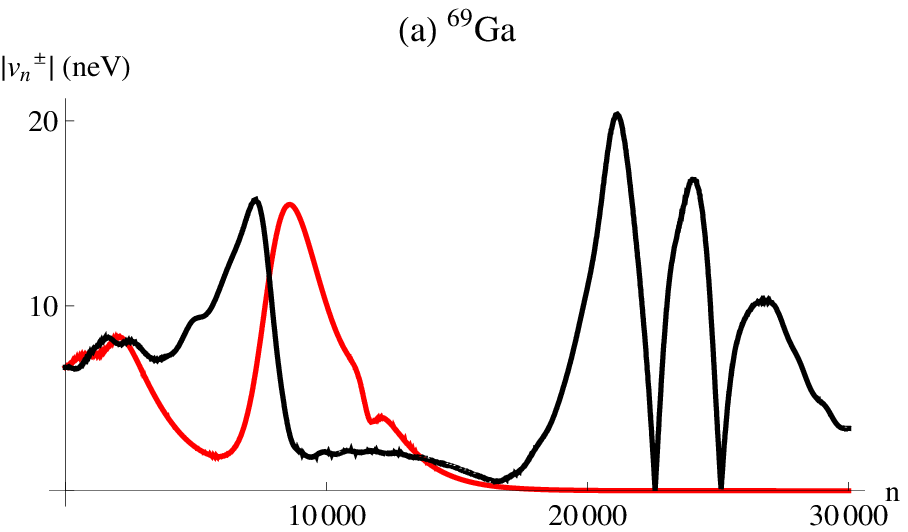}
} \\
{\includegraphics[width=0.9\columnwidth]{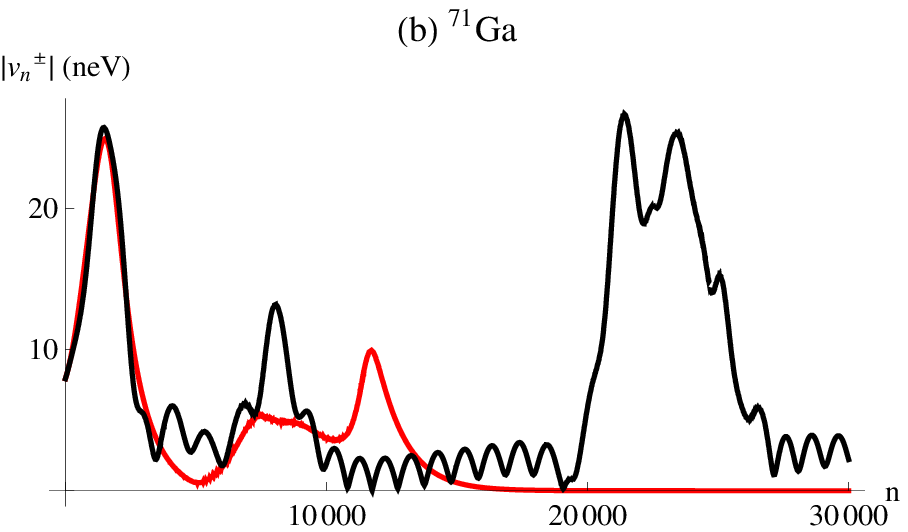}
} \\
{\includegraphics[width=0.9\columnwidth]{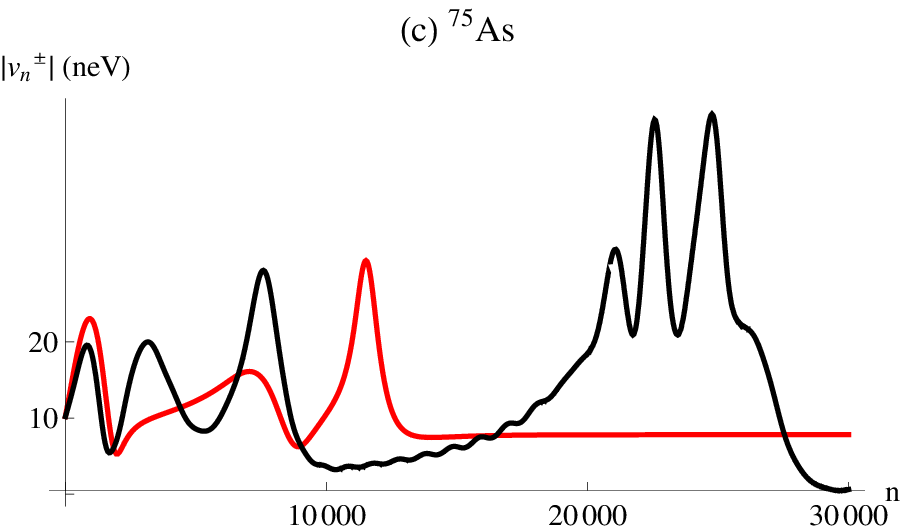}
} \\
\caption{ Transverse nuclear polarization as a function of sweep number $n$ for a GaAs double quantum dot with $v_\text{SO}=$8 neV. The duration of LZ pulses $T_\text{LZ}=80$ ns. Red curves show  results for the waiting time $T_w$ in resonance with the precession time of $^{75}$As, $T_w=t_{^{75}{\rm As}}$, while  black curves the data for 1\% off-resonance regime.
}
\label{fig:GaAs_SOresandnonresspecies}
\end{figure}

More insights into the sensitivity of the screening of SO coupling to the deviation from the resonance can be gained from Fig.~\ref{fig:GaAs_SOresandnonresspecies} that displays the contributions to $v_n^\pm$ from each of the species. Using the same intial conditions as in Fig.~\ref{fig:GaAs_SOresandnonrestot}, we plot  the evolution of the matrix elements $|v_n^\pm|$ for both the resonant and slightly off-resonance regimes. Initially they follow each other closely. However, after a couple of thousand sweeps, the deviations become significant. Ultimately, the contributions from $^{69}$Ga and $^{71}$Ga do not vanish in the non-resonant case, and the contributions from $^{75}$As does not screen the SO coupling.

The critical sensitivity of stroboscopic self-quenching to small deviations from resonance looks indicative of a chaotic behavior of the system.\cite{Lorenz} This is not surprising because the system of integro-differential equations of Eq.~(\ref{dynam}) is highly nonlinear because the coefficients $\mbox{\boldmath$\Delta$}_{j\lambda}$ depend through Eqs.~(\ref{eq11}) on the electronic amplitudes $c_S,c_{T_+}$ that, in turn, depend on all nuclear angular momenta ${\bf I}_{j\lambda}$. In this context, we speculate that a strong revival of all black curves in Fig.~\ref{fig:GaAs_SOresandnonresspecies} near $n\approx15000$ where all red curves saturate, and the return of black curves close to their initial values near $n\approx28000$, is reminiscent of the strange attractor pattern.\cite{Lorenz} These signatures of chaotic nuclear dynamic in SO coupled systems require a more detailed study.

We conclude that stroboscopic screening of the SO coupling is not a robust phenomenon. 

While the above simulations are focused on the large $n$ region, we mention that commensurability oscillations in the polarization accumulation per sweep were observed experimentally\cite{Foletti:2009} and described theoretically\cite{neder:2013} in the small $n$ region, $n\alt10^4$.

In addition to the regular investigation of the transverse magnetization, we also followed the time dependence of the longitudinal magnetization $v_n^z=V_{s}\sum_{\lambda}A_{\lambda}\sum_{j\in\lambda}\rho_{j\lambda}I^z_{j\lambda}$. In presence of SO coupling, it shows an oscillating sign-alternating behavior, and we were unable to detect any signatures of its accumulation. 

Summarizing the results of Secs.~\ref{InAs} and \ref{sec:GaAs}, we conclude that SO coupling eliminates self-quenching and causes the nuclei of a pumped system to exhibit a persistent irregular dynamics. We speculate that this phenomenon is closely related to the feedback loop technique for building controllable nuclear gradients which is inherently based on employing such a dynamics.\cite{Bluhm:prl10} Indeed, any long-term control of the nuclear ensemble by alternating $S\rightarrow T_+$ and $T_+\rightarrow S$ sweeps is impossible after the self-quenching set-in time that is of a millisecond scale in absence of SO coupling. Our data, especially Fig.~\ref{fig:GaAs_SOresandnonrestot}, suggest that near the resonance between the waiting time of LZ pulses and the Larmor frequency of one of the nuclear species the quasi-periods of nuclear fluctuations become longer and are controlled by the deviations from the exact resonance. We also expect that under these conditions the nuclear gradients should be dominated by the resonant specie.

\section{Conclusions}
\label{sec:conclude}

An analytical solution of a simplified model, and extensive numerical simulations for a realistic geometry, prove that self-quenching is a generic property of the central spin-$1/2$ problem in absence of spin-orbit coupling. As applied to a double quantum dot of a GaAs type, where electron and nuclear spins are coupled {\it via} hyperfine interaction, pumping nuclear magnetization across a $S$-$T_+$ avoided crossing through successive Landau-Zener sweeps  ceases after about $10^4$ sweeps. This is a result of the screening of the initial fluctuation of the nuclear magnetization by the injected magnetization and vanishing of the $S$-$T_+$ anticrossing width, and this sort of self-quenching is robust. Under the influence of moderate noise, the system wanders through a set of dark states belonging to a wast landscape of the system including about $10^6$ nuclear spins coupled through inhomogeneous electron spin density. With time intervals depending on the level of the noise, the system experiences revivals when additional magnetization is injected, and afterwards it wanders through a new set of dark states.

Due to the violation of the angular momentum conservation, spin-orbit coupling changes the situation drastically. Self-quenching sets in only stroboscopically under the condition that the waiting time between consecutive Landau-Zener sweeps is in resonance with the Larmor precession time of one of the nuclear species. Then the precessing Overhauser field of the resonant specie compensates the spin-orbit field $v_{\rm SO}$ during the sweep, while contributions of other species vanish. This sort of self-quenching is fragile and sensitive even to minor deviation from the resonance. Generically, injection of nuclear polarization oscillates in time and changes sign. Therefore, spin-orbit coupling causes the 
nuclear magnetization of a pumped $S$-$T_+$ double quantum dot to exhibit a persistent dynamics.

We suggest that the feedback loop technique for building controllable nuclear field gradients\cite{Bluhm:prl10} is based on the oscillatory behavior of the nuclear spin magnetization caused by spin-orbit coupling. Spin-orbit coupling is a natural mechanism of overcoming self-quenching. The technique employs persistent oscillations and selects the sign of the pumping response to the changing magnetization gradient.

\acknowledgments

A. B. would like to thank B. I. Halperin for his hospitality at Harvard University where this work was initiated. We are grateful to B. I. Halperin, C. M. Marcus, L. S. Levitov, H. Bluhm, K. C. Nowack, M. Rudner, and L. M. K. Vandersypen for useful discussions. E. I. R. was supported by the Office of the Director of National Intelligence, Intelligence Advanced Research Projects Activity (IARPA), through the Army Research Office grant W911NF-12-1-0354 and by the NSF through the Materials Work Network program DMR-0908070.

\appendix

\section{Spin-orbit coupling}
\label{sec:SO}

The Rashba spin-orbit Hamiltonian is 
\begin{equation}
H_{so}=\alpha\sum_{n=1,2}\left[\sigma_{x}(n)k_{y}(n)-\sigma_{y}(n)k_{x}(n)\right], \label{Hso}
\end{equation}
where $\alpha$ is the strength of SO interaction, and $k_{x}(n)$ and $k_{y}(n)$ are the in-plane momenta for the electron $n$. The singlet and triplet states are $\Psi_{S}(1,2)=\psi_{S}(1,2)\chi_{S}(1,2)$
and $\Psi_{T_+}(1,2)=\psi_{T}(1,2)\chi_{T_+}(1,2)$, where the orbital components of the singlet and triplet wave functions have been defined in Sec.~\ref{sec:numerical} and the spin parts of the wave functions are $\chi_{S}(1,2)=\left(| \uparrow_{1} \rangle | \downarrow _{2} \rangle - |\downarrow _{1} \rangle \uparrow _{2} \rangle\right)/\sqrt{2}$ and $\chi_{T_+}(1,2)=|\uparrow _{1} \rangle | \uparrow_{2} \rangle $. In terms of the orbital wave functions, the SO induced $S$-$T_{+}$ coupling is then $v_\text{SO}^{+}=i\alpha\cos\nu\left\langle \right. \psi_{R}\left|k_{x}+ik_{y}\left|\psi_{L}\left\rangle \right.\right.\right.$. 
This matrix element can be estimated similarly  to Refs.~\onlinecite{Rudner:prb10a} and \onlinecite{Stepanenko}. It depends exponentially on the overlap between the wave functions of the dots, $\psi_L$ and $\psi_R$. Therefore, the SO coupling $v_\text{SO}^{+}$ can be tuned and strongly decreases with the interdot distance $d$. In InAs quantum wires the SO coupling parameter is around $\alpha\sim$10$^{-11}$eVm\cite{Hernandez:prb10}
corresponding to a spin precession length of $l_{\rm SO}=\hbar^{2}/(2m^{*}\alpha)$
of around 100 nanometers. With typical parameters of $d=100$ nm and $l=50$ nm and $\cos\nu=1/\sqrt{2}$, we find that $v_\text{SO}$ is around $4\times10^{-5}$ eV. This is about two orders of magnitude larger than the typical $S$-$T_{+}$ coupling $10^{-7}$ eV induced
by the hyperfine interaction, but decreases with $d$ exponentially. Theoretical estimates of $v_\text{SO}$ hold only with exponential accuracy. Pre-exponential factors are model dependent, and a somewhat different estimate was proposed in Ref.~\onlinecite{Fasth}.
In GaAs quantum dots,\cite{Nowack:science07} the SO coupling constant $\alpha$ is two orders of magnitude smaller than in InAs with $l_{\rm SO}\approx30$ $\mu$m, so that the SO coupling
may be comparable to the hyperfine induced coupling, and is usually considered as weaker than it. These estimates should be treated with caution since the SO coupling is not only a function of the material but is sample specific. 

For GaAs, an estimate of a typical fluctuation as $v_n^0\approx A/\sqrt{N}$ with $A$ from Table II and $N\approx10^6$ results in $v_n^0\approx100$ neV. Our estimates of $v_{\rm SO}$ of Ref.~\onlinecite{Brataas:prl12} gave $v_{\rm SO}\approx50$ neV, while the estimate of Ref.~\onlinecite{neder:2013} is $v_{\rm SO}\approx15$ neV.\cite{caption} Recently Shafiei et al.\cite{Shafiei} managed to resolve the SO and hyperfine components of electric dipole spin resonance (EDSR)\cite{RS91} in the same system, a GaAs double quantum dot. Their data suggest that both components were of a comparable magnitude, with the hyperfine component maybe somewhat stronger.  

Keeping in mind the uncertainty in the values of $v_{\rm SO}$, we carry out simulations both with SO coupling and without it.



\begin{references}
\bibitem{also} This equation can be also derived by changing the variables in Eq.~\ref{eq:n} so that $\gamma$ depends on $I^z$ instead of $n$ by using $d \gamma/d I^z=(d \gamma/dn) / (dI^z/dn)$ which results in $d\gamma/dI^z=-A_0^2I^z/\beta^2$. Because this equation is independent of the sweep number $n$ and the probability $P$, it can be readily integrated.
\bibitem{numcomment} Because our procedures are numerical, they can be applied to an arbitrary $t$ dependencies of $\epsilon_S-\epsilon_{T_+}$. We choose a linear dependence only to stay closer to the terms of the standard Landau-Zener theory.
\bibitem{Q} It is seen from Figs.~4 - 6 that SO interaction makes the coupled electron-nuclei dynamics highly volatile. Oscillations of $|v_n^\pm|$ are accompanied with strong shakeups. E.g., the shakeup factor $Q$ is typically $Q\sim15$ in the peaks of $|v_n^\pm|$ for the $v_{\rm SO}=62$ neV curve of Fig.~6.
\end{references}
\end{document}